\documentclass[preprint,aps,pra,showpacs,floatfix]{revtex4-1}
\usepackage{graphicx}
\usepackage{times}
\usepackage {euscript}
\usepackage{amsthm}
\usepackage{epsf}
\usepackage{epsfig}
\usepackage{bm}
\usepackage{bbm}
\begin{document}
\title{ Relativistic calculations of the ground state energies and the critical distances for 
one-electron homonuclear  quasi-molecules}
\author{D.~V. Mironova$^{1,2}$, I.~I. Tupitsyn$^{1}$, V.~M. Shabaev$^{1}$,
 and G.~Plunien$^{3}$}

\affiliation{
$^1$ Department of Physics, St. Petersburg State University,
Oulianovskaya 1, Petrodvorets, 198504 St. Petersburg, Russia \\
$^2$ITMO University, Kronverkskii ave 49, 197101, Saint Petersburg, Russia
$^3$Institut f\"ur Theoretische Physik, Technische Universit\"at Dresden,
Mommsenstra{\ss}e 13, D-01062 Dresden, Germany \\
}

\begin{abstract}
The ground-state energies
of one-electron homonuclear quasi-molecules for the nuclear charge
number in the range  $Z=1-100$ at the ``chemical'' distances 
$R= 2/Z$ (in a.u.) are calculated. The calculations are performed
for both  point- and extended-charge nucleus cases
using the Dirac-Fock-Sturm approach with the basis functions constructed from the
one-center Dirac-Sturm orbitals. The critical distances $R_{\rm cr}$, 
at which the ground-state level reaches the edge of the negative-energy Dirac continuum,
are calculated for homonuclear quasi-molecules in the range: $ 85\leq Z\leq 100$. 
It is found that in case of  U$_2^{183+}$ 
the critical distance  $ R_{\rm cr} = 38.42$ fm 
for the point-charge nuclei and $R_{\rm cr} = 34.72$ fm  for 
 extended nuclei.
\end{abstract}

\pacs{34.10.+x, 34.50.-s, 34.70.+e}
\maketitle

\section{Introduction}

A one-electron diatomic quasi-molecule represents the simplest
molecular system. 
Precise calculations of one-electron homonuclear quasi-molecules
are generally used for tests of various theoretical
methods developed for calculations of diatomic systems.    
Theoretical analysis of the electronic structure of a
one-electron quasi-molecular system consists in solving the  
one-electron two-center Sch\"odinger or Dirac equation. 

In the nonrelativistic case the three-dimensional two-center Sch\"odinger 
equation can be transformed into two  
ordinary (one-dimensional) differential equations \cite{Burrau_27}
and, therefore, can be solved to a high accuracy \cite{Wind_65}. 
Moreover, the scaling $r'=r/Z$ allows one to reduce the solution
of the Scr\"odinger equation with the internuclear distance $R$ and the nuclear charge
$Z$ to the solution of the same equation for the molecular ion $H^{+}_{2}$ with
  the internuclear distance $R/Z$. This makes the molecular ion  $H^{+}_{2}$ to
be a good test system for various theoretical methods.
In the relativistic case, however, the variables can not be completely separated
and the simple scaling is no longer valid.
To date, various theoretical methods 
were developed to calculate  homonuclear 
quasi-molecules \cite{Roothaan_51, Laaksonen_84,
Laaksonen_86, Schulze_85, Heinemann_88, Heinemann_90,Yang_91,Kullie_01}.
Systematic calculations of the ground-states energies of
molecular ions for a wide range of $Z$ 
at the distances $R=2/Z$ were  
performed  in Ref. \cite{Sundholm_94}.

Investigations of quasi-molecules 
formed during low-energy heavy-ion collisions with the total nuclear charge 
larger than the critical value, 
$Z_1 +Z_2 \geq Z_{\rm cr}\approx 172$, can provide a unique possibility to study
quantum electrodynamics (QED) at supercritical electromagnetic fields
\cite{Zeldovich_71,Greiner_85}.  
It is known that the ground-state level reaches the edge 
of the negative-energy spectrum, when the internuclear distance 
$R$ becomes equal to the critical value $R_{\rm cr}$. For the distances 
$R < R_{\rm cr}$, the ground-state level dives into negative-energy 
Dirac continuum as a resonance. The critical distances $R_{\rm cr}$ were 
calculated for the point-charge nuclei 
in Refs. \cite{Rafelski_76, Lisin_77, Matveev_00} and for extended nuclei 
in Refs. \cite{Muller_76, Lisin_80, Popov_01}. 
However, since the first calculations for extended nuclei were accomplished 
using either a crude numerical approach  \cite{Muller_76} or 
an approximate analytical method  \cite{Lisin_80, Popov_01}, 
their accuracy was rather low. In case of  U$^{183+}_2$,
the most precise calculations 
of the  critical distance were performed in  
Refs. \cite {Tupitsyn_10,Artemyev_10}.

In the present work, high-precision relativistic calculations of
the ground-state energies of molecular ions with the nuclear charges 
in the range $Z=1-100$  
at ``chemical distances'' $R=2/Z$ (in a.u.) are performed.
We also calculate the critical distances $R_{\rm cr}$ for
one-electron quasi-molecules in the range: $ 85\leq Z\leq 100$.
All the calculations, being performed for both point- and 
extended-charge nuclei, are based on the Dirac-Fock-Sturm method 
\cite{Brattsev_77,Tupitsyn_03,Tupitsyn_05,Tupitsyn_10, Tupitsyn_12}. 
The basic equations of this method for the one-electron two-center problem
are given  in section \ref{sec: Theory}.
In section \ref{sec: Results},  we present the numerical results 
 and compare them with the calculations performed by other methods. 

Atomic units are used throughout the paper ($\hbar=m=e=1$).

\section{Dirac-Sturm method for the two-center problem} 
\label{sec: Theory}

In the framework of the Born-Oppenheimer approximation the 
electronic wave function  $\psi(\vec{r})$ is determined 
by the Dirac equation: 
\begin{equation}
\hat h_{\rm D} \psi_n(\vec{r}) = \varepsilon_n \, \psi_n(\vec{r}) \,,
\end{equation}
where $\varepsilon_n$ is the energy of the stationary state and
 $\hat h_{\rm D}$ is the two-center Dirac Hamiltonian defined by
\begin{equation}
\hat h_{\rm D} =c (\vec{\alpha} \cdot \vec{p}) \,+\, 
\beta \, c^2 \,+\, V_{AB}(\vec{r})\,.
\end{equation}
Here $c$ is the speed of light, $\vec{\alpha}$, $\beta$ are the Dirac matrices,
 $V_{AB}(\vec{r})$ is the two-center Coulomb potential,
\begin{equation}
V_{AB}(\vec{r}) = V_{\rm nucl}^{A}(\vec{r}_A) + V_{\rm nucl}^{B}(\vec{r}_B)
\,, \qquad \vec{r}_A=\vec{r}- \vec{R}_A\,, \qquad \vec{r}_B=\vec{r}- \vec{R}_B\,,
\end{equation}
and $\vec {R_{A}}$ and $\vec{R_{B}}$ determine the positions of the nuclei.
The one-center Coulomb potential:
\begin{equation}
V_{\rm nucl}(\vec{r}) = \left  \{ 
\begin{array}{ll}  \displaystyle
-Z/r &  \quad \hbox{for the point-charge nucleus\,,}
\\[2mm] \displaystyle
-\int d \vec{r}^{\prime} \, 
\frac{Z\rho_{\rm nucl}(\vec{r}^{\prime})}{|\vec{r}-\vec{r}^{\prime}|} &
 \quad \hbox{for the extended nucleus\,, }
\end{array}
\right .
\end{equation}
where the nuclear charge density $\rho_{\rm nucl}(\vec{r})$ is normalized to unity
($\int d\,\vec{r} \rho_{\rm nucl}(\vec{r})=1$).  
 
The two-center expansion of the stationary wave function $\psi_{n}(\vec{r})$ is given by
\begin{equation}
\psi_n(\vec{r})  =  \displaystyle \sum_{\alpha=A,B} \, \sum_{a}
c^{n}_{\alpha a} \, \varphi_{\alpha,a} (\vec{r}-\vec{R}_\alpha)\,,
\label{expan1}
\end{equation}
where index $\alpha=A,B$ labels the centers and index $a$
numerates the basis functions at the given center.
The coefficients $c^{n}_{a \alpha}$ of the expansion (\ref{expan1}) 
 can be obtained 
solving the generalized eigenvalue problem:
\begin{equation}
\sum_{k} H_{jk} \,c^{n}_{k} =  \displaystyle \varepsilon_n \,
\sum_{k} S_{jk} \,c^{n}_{k} \,,
\end{equation}
where subscripts $j$ and $k$ numerate the basis functions of both 
centers, and the matrix elements $H_{jk}$ and $S_{jk}$ are given by
\begin{equation}
H_{jk} \,=\, \langle j \mid \hat h_{\rm D} \mid k \rangle \,, \qquad
S_{j k} \,=\, \langle j \mid k \rangle \,.
\label{matr1}
\end{equation}

As the basic functions, we consider the central-field bispinors centered at the positions of the ions:
\begin{equation}
\varphi_{n\kappa m}(\vec{r}) =
\left  ( \begin{array}{l} \displaystyle
\,\, \frac{~P_{n \kappa}(r)}{r} \,  \chi_{\kappa m}(\Omega,\sigma)
\\[4mm] \displaystyle
i \, \frac{Q_{n \kappa}(r)}{r} \, \chi_{-\kappa m}(\Omega,\sigma)
\end{array} \right ) \,,
\end{equation}
where $P_{n \kappa}(r)$ and $Q_{n \kappa}(r)$ are the large and small radial
components, respectively, and $\kappa=(-1)^{l+j+1/2}(j+1/2)$ is the
relativistic angular quantum number.
The radial components are numerical solutions of the radial 
Dirac-Sturm equations in the central field potential $V(r)$:  
\begin{equation}
\begin{array}{lll}  \displaystyle
c \left (-\frac{d}{dr}+\frac{\kappa}{r} \right ) \,\overline Q_{n\kappa}(r) +
\left (V(r)+c^2 - \varepsilon_{n_0\kappa} \right) \,
\overline P_{n\kappa}(r) &=&
\lambda_{n \kappa} \, W_{\kappa}(r) \, \overline P_{n \kappa}(r)\,,
\\[3mm] \displaystyle
c \left(~~\frac{d}{dr}+\frac{\kappa}{r} \right) \, \overline P_{n \kappa}(r) +
\left (V(r)-c^2  - \varepsilon_{n_0 \kappa} \right) \,
\overline Q_{n \kappa}(r) &=&
\lambda_{n \kappa} \, W_{\kappa}(r) \, \overline Q_{n \kappa}(r)\,.
\end{array}
%
\label{sturm1}
\end{equation}
Here $\lambda_{n \kappa}$ can be considered as an eigenvalue of the
Dirac-Sturm operator and $W_{\kappa}(r)$ is a constant sign weight function.
In our calculations we use the following weight function:
\begin{equation}
W_{\kappa}(r)  \,=\, - \, \frac{1 \,-\, 
\exp(-(\alpha_{\kappa} \, r)^2)}{(\alpha_{\kappa} \, r)^2}\,.
\label{sturm2}
\end{equation}
In contrast to $1/r$, this weight function is regular at the origin.
The Sturmian operator is Hermitian and does not contain any continuum
spectra. Therefore, the generalized eigenvalue equation with the weight 
function (\ref{sturm2}) yields a complete and discrete set of eigenfunctions
that are orthogonal to each other with the weight (\ref{sturm2}).

Equations (\ref{sturm1}) are solved using the finite difference method 
with a constant step on Brattsev's grid 
$\rho=\alpha \, r +\beta \ln(r)$ \cite{Brattsev_77}.
These solutions, which  have the right asymptotic 
behavior at the origin and infinity, are used to construct
the basis set. 
In particular, for the two Coulomb point-charge centers 
the behavior of the basic functions at the origin is characterized by 
the fractional
degree of the radius, $\sim r^{\gamma}$ with $\gamma=\sqrt{\kappa^2 - (Z/c)^2}$.

The central-field potential $V(r)$ in equations (\ref{sturm1}) can be 
chosen to provide the most appropriate basis. For instance,
at small internuclear distances the potential
$V(r)$ at the center $A$, 
in addition to 
the Coulomb potential of the nucleus $A$,
 $ V^{A}_{\rm nucl}(r)$,
should also include  the monopole part  of the
reexpansion  of the potential
$V^{B}_{\rm nucl}(\vec{r}-\vec{R_B})$ with respect to the center $A$:
\begin{equation}
\label{v}
V^{A}(r) = V^{A}_{\rm nucl}(r) + V^{B}_{\rm mon}(r) \,,
\end{equation}
where 
\begin{equation}
V^{B}_{\rm mon}(r) = \frac{1}{4 \pi} \,\int d\Omega_A \,\,
V^{B}_{\rm nucl}(\vec{r}-\vec{R_B}) \,. 
\end{equation}
However, for the ``chemical'' distances ($R=2/Z$)  
taking into account
the monopole potential of the second ion in Eq. (\ref{v})
does not improve the convergence of the results 
with respect to the number 
of the basis functions. For this reason, we keep this term
evaluating the critical distances and
neglect it in the calculations at the ``chemical'' distances.
%
\section{Results and discussion}
\label{sec: Results}

High-precision relativistic calculations of the  $1\sigma_{g}$ state energy 
of one-electron homonuclear quasi-molecules at the distance $R=2/Z$ (in a.u.)
have been performed employing  
the Dirac-Sturm method. The results of these calculations
for the point-  and extended-charge nuclei are given 
in Table \ref{all_Z}.  
The extended-nucleus results were obtained  using the Fermi 
model of the nuclear charge distribution:
\begin{eqnarray}\label{fermi}
\rho_{\rm nucl}(r) = \frac{N}{1+\exp{[(r-r_0)/a]}}\,,
\end{eqnarray}
where the parameter $a$ was chosen to be $a=2.3/(4\ln{3})$ 
and the parameters $N$ and $r_0$ are obtained using the values
of  the root-mean-square (rms) nuclear charge radii 
$\langle r^2_{\rm n} \rangle^{1/2}$ taken from Refs. 
\cite{Angeli_13,Johnson_85}. 
The point-nucleus results were recently
presented in Ref. \cite{Tupitsyn_14}.
 In these calculations we used 
the speed of light as obtained from
the fine structure constant $\alpha = 1/c$ (the value of $\alpha$
is taken from CODATA \cite{CODATA_10}).

\begin{table}
\small
\begin{center}
\vspace{1mm}
\caption{Relativistic energies (a.u.) of the 1$\sigma_{g}$ quasi-molecular 
state for the point- and extended-charge nuclei and  $R=2/Z$ a.u. 
(speed of light c = 137.035999074 \cite{CODATA_10})}
\label{all_Z}
\vspace{2mm}
\begin{tabular}{c|c|c|c|c}
\hline \hline
&&&& \\[-3mm]
                 \hspace{1mm} Z        \hspace{3mm}
               & \hspace{3mm} Ion      \hspace{2mm}
               & \hspace{11mm} $\varepsilon_{\rm 1\sigma_g}$ ({\rm point-charge nucl.})    \hspace{10mm}
               & \hspace{3mm} 
                 $\varepsilon_{\rm 1\sigma_g}$ ({\rm extended-charge  nucl.})
                 \hspace{1mm} \\[2mm]
\hline \hline
&&&& \\[-3mm]
1  & $\rm H_2^{+}$~~~  &  -1.102641581032  &  \\[-0mm]
2  & $\rm He_2^{3+}$~  &  -4.410654728260  & -4.410654714140 \\[-0mm]
10 & $\rm Ne_2^{19+}$  &  -110.3372043998  & -110.3371741499 \\[-0mm]
20 & $\rm Ca_2^{39+}$  &  -442.2399970985  & -442.2392996469 \\[-0mm]

30 & $\rm Zn_2^{59+}$  &  -998.4267621737  & -998.4214646525 \\[-0mm]
40 & $\rm Zr_2^{79+}$  &  -1783.587352661  & -1783.563450815 \\[-0mm]
50 & $\rm Sn_2^{99+}$  &  -2804.659800901  & -2804.571434254 \\[-0mm]
60 & $\rm Nd_2^{119+}$ &  -4071.309814908  & -4071.036267926 \\[-0mm]
70 & $\rm Yb_2^{139+}$ &  -5596.754834761  & -5595.926978290 \\[-0mm]
80 & $\rm Hg_2^{159+}$ &  -7399.228750561  & -7397.028800116 \\[-0mm]
90 & $\rm Th_2^{179+}$ &  -9504.756648531  & -9498.588788490 \\[-0mm]
92 & $\rm U_2^{183+}$  &  -9965.365357898  & -9957.775519122 \\[-0mm]
100 & $\rm Fm_2^{199+}$ & -11952.94176701  & -11936.41770218 \\[-0mm]
\hline \hline
\end{tabular}
\end{center}
\label{Ground_energies}
\end{table} 


In Table \ref{Comparison}, to demonstrate  the accuracy of our approach,
we  compare the point-nucleus results for 
the ground-state energies 
of the molecular ions H$_2^{+}$, Th$_2^{179+}$, and U$_2^{179+}$ at
the internuclear distances $R=2/Z$ obtained with different methods.
In this table the value of the speed of light is chosen to be  $c= 137.0359895$, as in 
our previous work \cite{Tupitsyn_14}. As one can see from the table, our results  
 \cite{Tupitsyn_14}
are in a good agreement with the previous calculations reported 
in the literature. 
We also present the results for the nonrelativistic ground-state 
energy of the molecular ion H$_2^+$.  In our work, this result was obtained
by performing the calculation with the light speed 
$c_{\infty}=c \cdot 10^6$. Our value is in perfect agreement
with the most precise nonrelativistic calculation of Refs. 
\cite{Wind_65, Ishikawa_12}. 

\begin{table}[hbt]
\small
\caption{Comparison of the relativistic and non-relativistic ($c = \infty$) 
ground-state energies (in a.u.) 
of one-electron molecular ions at $R=2/Z$ (a.u.) for the point-charge nuclei
with other data 
reported in the literature.}
\label{Comparison}
\vspace{1.0mm}
\begin{center}
\vspace{-2.0mm}
\begin{tabular}{ c|l|l|l|l }
\hline\hline
&&& \\[-6mm]
\hspace{26mm} & \hspace{6mm} $c$ (a.u.) \hspace{4mm} & ~~~~~~~
H$_2^+$ $(Z=1)$~~~~~~ &~~Th$_2^{179+}$ $(Z=90)$~~ &
~~ U$_2^{183+}$  $(Z=92)$ \\[0mm]
\hline\hline
&&&\\[-5.5mm]

Our result \cite{Tupitsyn_14}      &~ $c= \infty$~ & ~ -1.102634214494   &    & \\[-1mm]
Wind \cite{Wind_65},
Ishikawa {\it et al.} \cite{Ishikawa_12}   &~ $c= \infty$  & ~ -1.1026342144949  &    & \\[-1mm]
Our result   \cite{Tupitsyn_14}     &~ 137.0359895~&~ -1.1026415810330  &~ -9504.756746927 
                        &~ -9965.365468058\\[-1mm]
Kullie and Kolb \cite{Kullie_01}      &~ 137.0359895 &~ -1.10264158103358 &~ -9504.756746923
                        & ~~~~~~~~~~ --  \\[-1mm]
Yang {\it et al.} \cite{Yang_91}       &~ 137.0359895 &~ -1.1026415810336  &~~~~~~~~~ ---
                        & ~~~~~~~~~~ --  \\[-1mm]
Ishikawa {\it et al.} \cite{Ishikawa_08}   &~ 137.0359895  & ~ -1.102641581033  &~~~~~~~~~ ---
                        & ~~~~~~~~~~ --  \\[-1mm]                        
Ishikawa {\it et al.} \cite{Ishikawa_12}   &~ 137.0359895  & ~ -1.102641581026  &~~~~~~~~~ ---
                        & ~~~~~~~~~~ --  \\[-1mm]                        
Parpia and Mohanty \cite{Parpia_95}   &~ 137.0359895 &~ -1.1026415801     &~ -9504.756696
                        & ~~~~~~~~~~ --  \\[-1mm]
Artemyev {\it et al.} \cite{Artemyev_10}    &~ 137.036     &~ -1.1026409        &~ -9504.752
                        &~  -9965.375    \\[-1mm]
Tupitsyn  {\it et al.}  \cite{Tupitsyn_10}     &~ 137.036     & ~ -1.1026405       &~ -9504.732
                        &~  -9965.307    \\[-1mm]
Alexander and Coldwell \cite{Alexander_99}    &~ 137.03606   &~ -1.102565         &~ -9498.98
                        & ~~~~~~~~~~ --  \\[-1mm]
Sundholm \cite{Sundholm_94}    &~ 137.03599    &~ -1.102641581      &~ -9461.9833
                        & ~~~~~~~~~~ --  \\[ 1mm]
\hline\hline
\multicolumn{5}{l}{}\\[-6mm]
\multicolumn{5}{l}{} 
\\[-6mm]
\end{tabular}
\end{center}
\label{Comparison}
\end{table}

\begin{figure}
\centering
\includegraphics[width=11cm,clip,angle=-90]{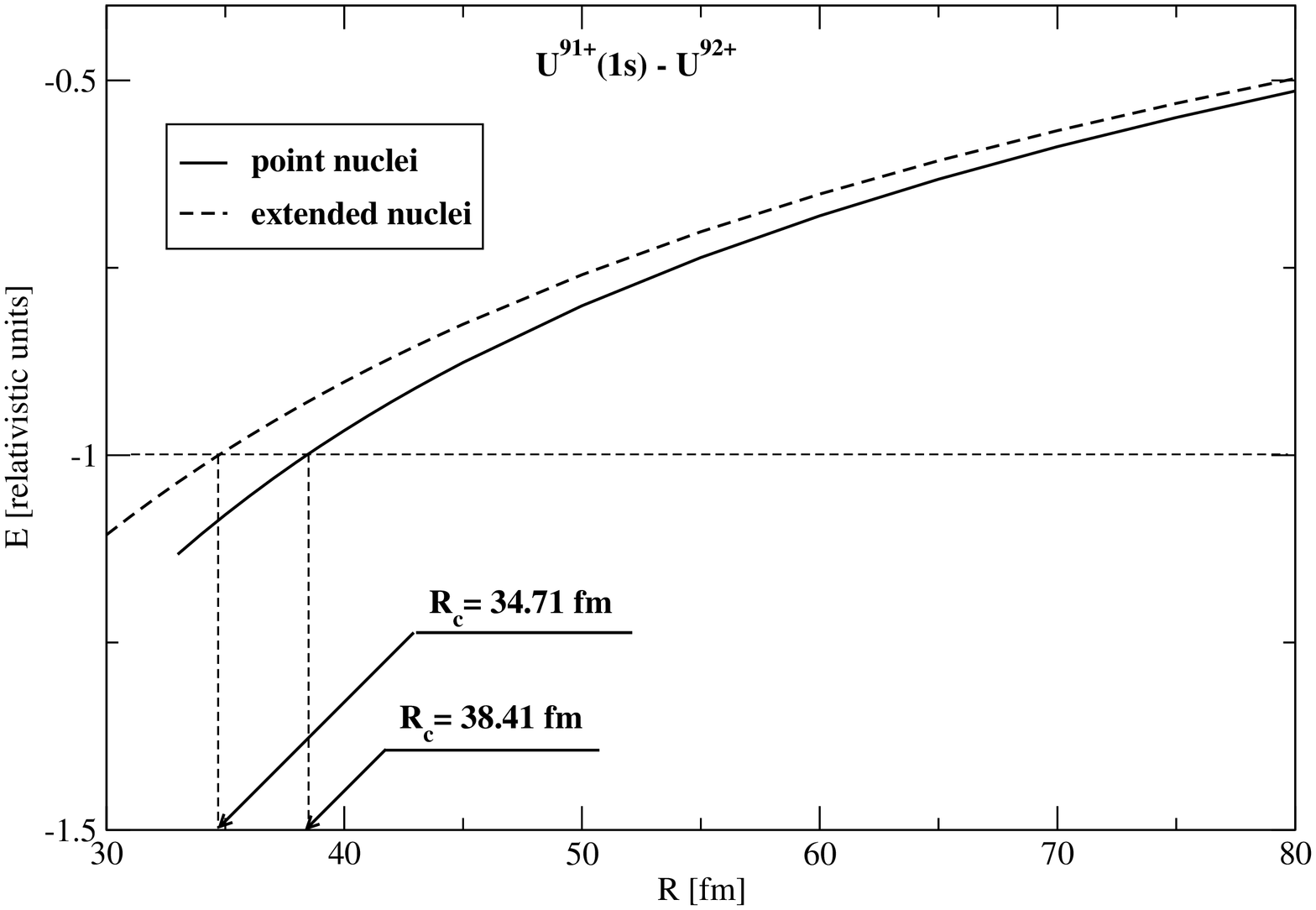}
\vspace{-6mm}
\caption{\small The $1\sigma_g$ state energy of the U$_2^{183+}$ quasi-molecule as a
function of the internuclear distance $R$ on a logarithmic scale.}
\label{U2_energies_1}
\end{figure}

\begin{figure}
\centering
\includegraphics[width=11cm,clip,angle=-90]{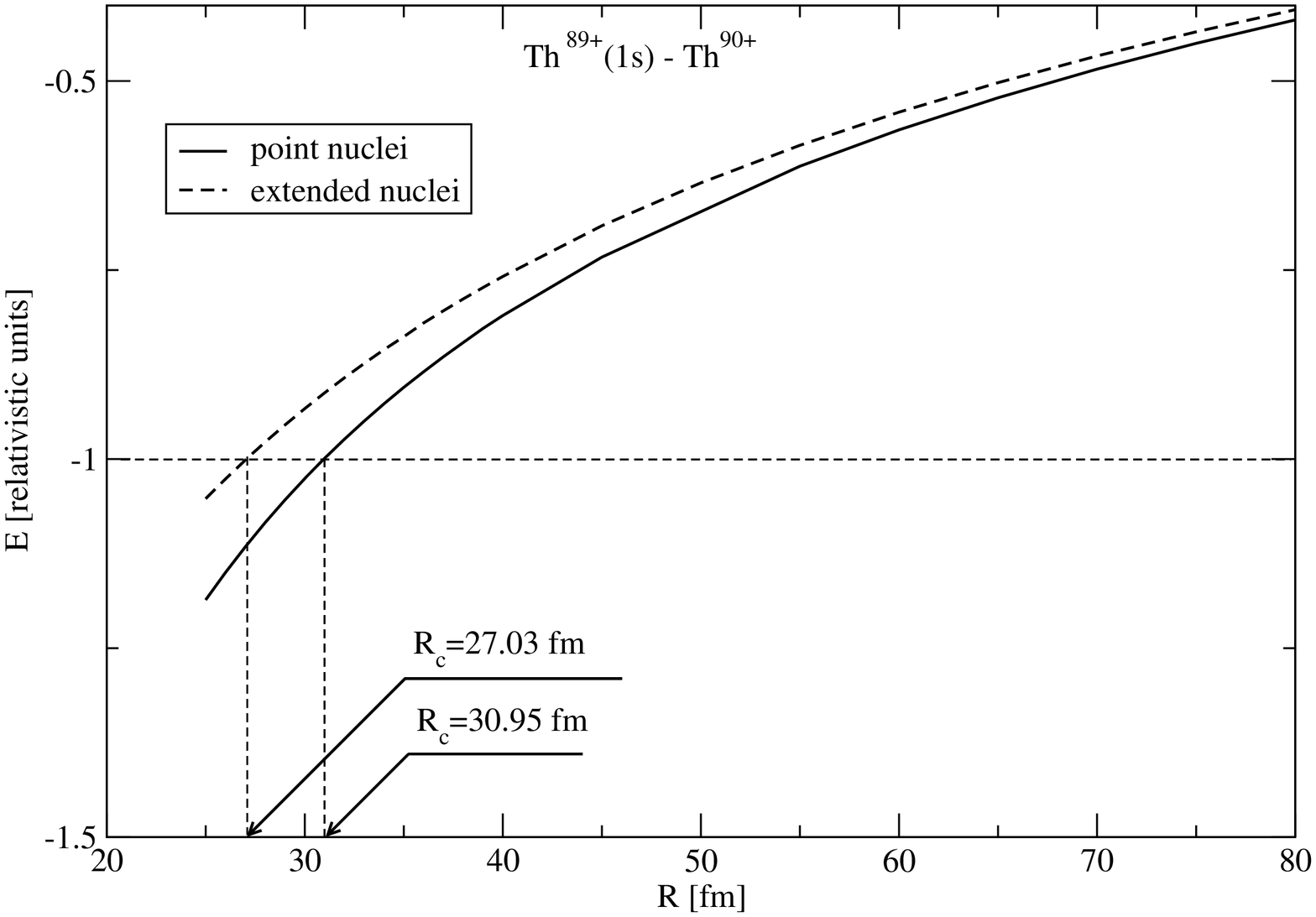}
\vspace{-6mm}
\caption{\small The $1\sigma_g$ state energy of the Th$_2^{179+}$ quasi-molecule as a
function of the internuclear distance $R$ on a logarithmic scale.}
\label{Th2_energies_1}
\end{figure}

In Fig. \ref{U2_energies_1} we display the energy of 
the $1\sigma_g$ state of the U$_2^{183+}$ quasi-molecule as a function
of the internuclear distance $R$ on a logarithmic scale.
In this figure the solid line indicates the energy
$E(R)$ calculated for the point-charge nuclei. The dashed line represents 
the related results for the extended-charge  nuclei,
which were obtained for the Fermi model (\ref{fermi}).
As one can see from Fig. \ref{U2_energies_1},
the 1$\sigma_g$ level dives into the negative-energy 
Dirac continuum at the critical distance $R_{\rm cr} = 38.42$ fm 
for the point-charge nuclei and at $R_{\rm cr} = 34.72$ fm for the extended-charge nuclei. 
In Fig. \ref{Th2_energies_1} we display the corresponding results for 
the Th$_2^{179+}$ quasi-molecule. In this case we observe that the ``diving'' point 
 occurs at the critical distance 
$R_{\rm cr} = 30.96$ fm for the point-charge nuclei and at $R_{\rm cr} = 26.96$ 
fm for the extended-charge nuclei. 


In  Table \ref{Crit_dist} we present the results of our two-center 
calculations of the critical distances $R_{\rm cr}$ for one-electron homonuclear
quasi-molecules  A$_2^{(2Z-1)+}$ for the point- and extended-charge nuclei and 
compare them with the previous calculations. 
It can be seen that our results for the point-charge nuclei are in a very good 
agreement with the results of Refs. \cite{Lisin_77,Tupitsyn_10}. 
As to the extended-nucleus case,
we can systematically compare our results only with the data 
obtained in Ref. \cite{Lisin_80}. The discrepancy between our data and those 
from Ref. \cite{Lisin_80} is considerably larger for the extended-nucleus case 
than for the point-nucleus case. We think this is due to 
a rather crude analytical
estimate of the nuclear size effect in Ref. \cite{Lisin_80}. 

\begin{table}[hbt]
\small
\caption{Critical distances $R_{\rm cr}$ (fm) for one-electron homonuclear
quasi-molecules A$_2^{(2Z-1)+}$.}
\label{Crit_dist}
\begin{center}
\vspace{1.0mm}
\begin{tabular}{ c|c|c|c|c|c  }
\hline \hline
\multicolumn{1}{c|}{} & \multicolumn{2}{|c|}{} & \multicolumn{3}{|c }{}\\[-7mm]
\multicolumn{1}{c|}{} & \multicolumn{2}{|c|}{Point nucleus} &
\multicolumn{3}{|c }{Extended nucleus}\\[2mm]
\hline  &&&&& \\[-7mm]
~~~$Z$~~~ &\quad ~This work~ \quad & \quad ~~~Others~~~\quad & \quad $\langle r^2_{\rm n} \rangle^{1/2}$(fm)
\quad  &\quad ~This work~ \quad &\quad ~Others~\quad \\[2mm]
\hline \hline  &&&&&\\[-5.5mm]
 85  & 15.61 &  15.61$^d$                                         \\[-2mm]
 86  & 18.29 &  18.29$^d$    & 5.5915$^f$ & 12.86 & 12.7$^{c}$    \\[-2mm]
 87  & 21.16 &  21.16$^d$    & 5.5915$^f$ & 16.42 & 16.0$^{c}$    \\[-2mm]
 88  & 24.24 &  24.24$^a$    & 5.6079$^f$ & 19.89 & 19.4$^{c}$    \\[-2mm]
     &       &  24.27$^d$    &            &       & 19.91$^{d}$   \\[-1mm]
 89  & 27.51 &  27.51$^a$    & 5.6700$^g$ & 23.38 & 22.9$^{c}$    \\[-1mm]
 90  & 30.96 &  30.96$^a$    & 5.7848$^f$ & 26.96 & 26.5$^{c}$    \\[-2mm]
     &       &  30.96$^d$    &            &       & 27.05$^{d}$   \\[-1mm]            
 91  & 34.60 &  34.60$^a$    & 5.7000$^g$ & 30.90 & 30.3$^{c}$    \\[-1mm]
 92  & 38.42 &  38.42$^a$    & 5.8571$^f$ & 34.72 & 34.3$^{c}$    \\[-2mm]
     &       &  36.8$^{~b}$~ &            &       & 34.7$^{e}$    \\[-2mm]
     &       &  38.43$^d$    &            &       & 34.72$^{d}$   \\[-1mm]
 93  & 42.41 &  42.41$^a$    & 5.7440$^g$ & 38.93 & 38.4$^{c}$    \\[-1mm]
 94  & 46.57 &  46.57$^a$    & 5.8601$^f$ & 43.10 & 42.6$^{c}$    \\[-2mm]
     &       &  46.58$^d$    &            &       & 43.16$^{d}$   \\[-1mm]
 95  & 50.89 &  50.89$^a$    & 5.9048$^f$ & 47.47 & 47.0$^{c}$    \\[-1mm]
 96  & 55.37 &  55.37$^a$    & 5.8429$^f$ & 52.06 & 51.6$^{c}$    \\[-2mm]
     &       &  55.58$^d$    &            &       & 52.09$^{d}$   \\[-1mm]
 97  & 60.01 &  60.01$^a$    & 5.8160$^g$ & 56.77 & 56.3$^{c}$    \\[-1mm]
 98  & 64.80 &  64.79$^a$    & 5.8440$^g$ & 61.56 & 61.0$^{c}$    \\[-2mm]
     &       &  64.79$^d$    &            &       & 61.63$^{d}$   \\[-1mm]
     &       &               &            &       & 61.1$^{e}$    \\[-1mm]
 99  & 69.73 &  69.73$^a$    & 5.8650$^g$ & 66.50 & 66.0$^{c}$    \\[-1mm]
 100 & 74.81 &  74.81$^a$    & 5.8860$^g$ & 71.57 & 71.1$^{c}$    \\[ 0mm]
\hline \hline 
\multicolumn{6}{l}{}\\[-6mm]
\multicolumn{6}{l}{$^a$ Ref. \cite{Lisin_77}, $^b$ Ref. \cite{Rafelski_76},
$^c$ Ref. \cite{Lisin_80}, $^d$ Ref. \cite{Tupitsyn_10}, $^e$ Ref. \cite{Muller_76},
$^f$ Ref.\cite{Angeli_13}, $^g$ Ref. \cite{Johnson_85}} 
\\[-6mm]
\end{tabular}
%
%
\end{center}
\label{Crit_dis}
\end{table}
\section{Conclusion}
In this work we applied the Dirac-Sturm method to calculate
the ground-state energies of one-electron homonuclear quasi-molecules 
with different 
nuclear charge numbers $Z$ at the 
internuclear distances $R = 2/Z$. The critical distances, at which
the ground state level of a heavy quasi-molecule reaches the edge of
the negative-energy Dirac continuum, were also calculated.
The calculations were performed for both point- and extended-charge
nuclei. As the result, the most precise data for the energies and the
critical distances are obtained. This also  demonstrates  high
efficiency of the Dirac-Fock-Sturm method in its application to
diatomic molecules.
   
\section{Acknowledgments}
This work was supported by RFBR (Grants No. 13-02-00630  and  No.11-02-00943), 
by  St.Petersburg State University (Grants No. 11.0.15.2010,
No. 11.38.261.2014, and No. 11.38.269.2014), and by DAAD.
%
%

\pagebreak
\clearpage
%


\begin{thebibliography}{99}
%
\bibitem{Burrau_27} D. Burrau, Kgl. Danske Videnskab Selskab.
                    Mat.~Fys.~Medd. {\bf 7}, No.14 (1927).
%
%
\bibitem{Wind_65} H.  Wind, J.~Chem.~Phys. {\bf 42}, 2371 (1965). 

%
\bibitem{Roothaan_51} C.C.J. Roothaan, Rev.~Mod.~Phys.{\bf 23}, 69 (1951). 
%
\bibitem{Laaksonen_84} L. Laaksonen and P. Grant, Chem.~Phys.~Letters. {\bf 5}, 485 (1984). 

%
\bibitem{Laaksonen_86} L. Laaksonen, P. Pyykk\"o, and D. Sundholm, Comput.~Phys.~Rept. {\bf 4}, 313 (1986).

%
\bibitem{Schulze_85} W. Schulze and D. Kolb,  Chem.~Phys.~Lett. {\bf 122}, 271 (1985).

%
\bibitem{Heinemann_88} D. Heinemann, B. Fricke, and D. Kolb, Phys.~Rev.~A. {\bf 38}, 4998 (1988).

%
\bibitem{Heinemann_90} D. Heinemann, A. Rosen, and B. Fricke, Physica Scripta. {\bf 42}, 692 (1990).
%
\bibitem{Yang_91}  L. Yang, D. Heinemann, and D. Kolb, Chem. Phys. Lett.
                   {\bf 178}, 213 (1991).
%
\bibitem{Kullie_01} O. Kullie, and D. Kolb, Eur. Phys. J. D {\bf 17}, 167 (2001).

%
\bibitem{Sundholm_94} D.~Sundholm, Chem.~Phys.~Lett. {\bf 223}, 469 (1994).

\bibitem{Zeldovich_71} Y.B. Zeldovich and V.S. Popov, Usp. Fiz. Nauk
                       {\bf 105}, 403 (1971)
                       [Sov. Phys. Usp. {\bf 14}, 673 (1972)].
\bibitem{Greiner_85} W. Greiner, B. M\"uller, and J. Rafelski,
Quantum Electrodynamics of Strong Fields (Springer-Verlag, Berlin,
1985).
%
\bibitem{Rafelski_76}  J. Rafelski and B. M\"uller, Phys. Lett.
                       {\bf 65}B, 205 (1976).
%
\bibitem{Lisin_77} V.I. Lisin, M.S. Marinov, and V.S. Popov, Phys. Lett.
                    {\bf 69}B, 2 (1977).
                    
%
\bibitem{Matveev_00} V.I. Matveev, D.U. Matrasulov, and H.Yu. Rakhimov,
                     Phys. At. Nucl. {\bf 63}, 318 (2000).

%
\bibitem{Muller_76} B. M\"uller and W. Greiner, Z. Naturforsch.
                    {\bf 31}a, 1 (1976).
%
\bibitem {Lisin_80}   V.~I.~Lisin, M.~S.~Marinov and V.~S.~Popov  
                      Phys. Letters, Vol.91B, 1 (1980)

%
\bibitem{Popov_01} V.S. Popov, Phys. At. Nucl. {\bf 64},
                   367 (2001).

%
\bibitem{Tupitsyn_10} I.~I.~Tupitsyn, Y.~S.~Kozhedub, V.~M.~Shabaev,
G.~B.~Deyneka, S.~Hagmann, C.~Kozhuharov G.~Plunien and Th.~St\"ohlker  
Phys. Rev. A 82, 042701 (2010)

%
\bibitem{Artemyev_10} A.N. Artemyev, A. Surzhykov, P. Indelicato, G. Plunien,
                      and Th. St\"{o}hlker,  J.~Phys.~B: At.~Mol.~Opt.~Phys. {\bf 43}, 235207 (2010).

%
\bibitem{Brattsev_77} V.F.~Brattsev, G.B.~Deyneka, and I.I.~Tupitsyn,
                     Bull.Acad.Sci. USSR, Phys. Ser. {\bf 41},
                     173 (1977).
%
%
\bibitem{Tupitsyn_03} I.I. Tupitsyn, V.M. Shabaev, J.R. Crespo L\'opez-Urrutia,
                      I. Draganic, R. Soria Orts, and J. Ulrich,
                      Phys. Rev. A {\bf 68}, 022511 (2003).
%
\bibitem{Tupitsyn_05} I.I. Tupitsyn, A.V. Volotka, D.A. Glazov, V.M. Shabaev,
                      G. Plunien, J.R. Crespo L\'opez-Urrutia, A. Lapierre,
                      and J. Ullrich,  Phys. Rev. A {\bf 72}, 062503 (2005).
%
%
%
%
%
\bibitem{Tupitsyn_12} I.I.~Tupitsyn, Y.S.~Kozhedub, V.M.~Shabaev,
       A.I.~Bondarev, G.B.~Deyneka, I.A.~Maltsev, S.~Hagmann, G.~Plunien,
       and Th.~Stohlker,  Phys.~Rev.~A {\bf 85}, 032712 (2012).

%
\bibitem{Angeli_13} I. Angeli and K.P. Marinova, At. Data Nucl. Data Tables {\bf 99},
                    69 (2013).
%
\bibitem{Johnson_85} W.R. Johnson  and G. Soff, At. Data Nucl. Data Tables {\bf 33},
                    405 (1985).                                 

%
\bibitem{Tupitsyn_14} I.I.~Tupitsyn and D.V.~Mironova, Optics and Spectroscopy, to be published.

%
\bibitem{CODATA_10} P.J. Mohr, B.N. Taylor, and D.B.Newell, 
                     Rev.Mod.Phys.,{\bf 84}, 1527 (2012).

%
\bibitem{Ishikawa_08} A. Ishikawa, H. Nakashima, and H. Nakatsuji, J. Chem.~Phys. {\bf 128}, 124103 (2008).
%
\bibitem{Ishikawa_12} A. Ishikawa, H. Nakashima, and H. Nakatsuji, Chem.~Phys. {\bf 401}, 62 (2012). 

%
\bibitem{Alexander_99} S.A.  Alexander and R.L. Coldwell, Phys.~Rev.~E {\bf 60}, 3374 (1999). 
%
\bibitem{Parpia_95} F.A. Parpia and A.K. Mohanty, Chem.~Phys.~Lett. {\bf 238}, 209 (1995).

%
                   
%
\end{thebibliography}
\end{document}